\pdfoutput=1

\documentclass[11pt,a4paper]{article}
\usepackage{times,latexsym}
\usepackage{url}
\usepackage[T1]{fontenc}
\usepackage{multirow}
\usepackage{booktabs,array,siunitx}
\usepackage{subcaption}

\usepackage{acl}

\usepackage{amsmath,amsfonts,bm}

\def\eqref#1{equation~\ref{#1}}

\def\1{\bm{1}}

\def\ve{{\bm{e}}}

\DeclareMathAlphabet{\mathsfit}{\encodingdefault}{\sfdefault}{m}{sl}
\SetMathAlphabet{\mathsfit}{bold}{\encodingdefault}{\sfdefault}{bx}{n}

\usepackage{times}
\usepackage{latexsym}
\usepackage{booktabs}

\usepackage[utf8]{inputenc}

\usepackage{microtype}

\usepackage{inconsolata}

\usepackage{hyperref}
\usepackage{url}
\usepackage{graphicx}
\usepackage{verbatim}
\usepackage{todonotes}
\usepackage{xspace}
\usepackage{multirow}
\usepackage{siunitx}
\usepackage{booktabs}
\usepackage[inline]{enumitem}

\usepackage{pifont}

\title{MAD Speech: Measures of Acoustic Diversity of Speech}

\author{Matthieu Futeral\thanks{\hspace{1mm} Work done while interning at Google DeepMind. Correspondence to \texttt{matthieu.futeral@inria.fr}} \\ Inria Paris \\ \And Andrea Agostinelli \\ Google DeepMind \And Marco Tagliasacchi \\ Google DeepMind \AND Neil Zeghidour \\ Kyutai \And Eugene Kharitonov\thanks{\hspace{1mm} Correspondence to \texttt{kharitonov@google.com}} \\ Google DeepMind
}

\begin{document}
\maketitle
\begin{abstract}

Generative spoken language models produce speech in a wide range of voices, prosody, and recording conditions, seemingly approaching the diversity of natural speech. However, the extent to which generated speech is acoustically diverse remains unclear due to a lack of appropriate metrics. We address this gap by developing lightweight metrics of acoustic diversity, which we collectively refer to as MAD Speech. We focus on measuring five facets of acoustic diversity: voice, gender, emotion, accent, and background noise.
We construct the metrics as a composition of specialized, per-facet embedding models and an aggregation function that measures diversity within the embedding space. Next, we build a series of datasets with a priori known diversity preferences for each facet. Using these datasets, we demonstrate that our proposed metrics achieve a stronger agreement with the ground-truth diversity than baselines. Finally, we showcase the applicability of our proposed metrics across several real-life evaluation scenarios. MAD Speech is made publicly available\footnote{\href{https://github.com/MatthieuFP/MAD_Speech}{https://github.com/MatthieuFP/MAD\_Speech}}.

\end{abstract}

\section{Introduction}

\looseness=-1
Recent progress in generative spoken language modeling~\citep{Lakhotia2021, Borsos2023,valle,spear-tts, soundstorm, audiopalm,voicebox,audiobox,Hassid2024,nguyen2024spirit} brought models that can generate speech in a wide range of voices, prosody and recording conditions as found in large-scale datasets of natural speech~\cite{Kahn2020,mls,common-voice}. 

\looseness-1These generative models are typically evaluated in terms of faithfulness to their inputs, i.e.\ speaker identity and transcript~\citep{Borsos2023,valle,spear-tts}. Thus it remains unknown how diverse their outputs are, since those metrics neither capture variability nor take into account factors of the perceived speech diversity (i.e., recording conditions, prosody, styles, and accents). Being able to measure speech diversity would also help detect mode collapse when finetuning models~\citep{kirk2023understanding}, selecting training \& inference hyperparameters, and optimizing for human feedback~\citep{musicrl}. Moreover, such measurements are vital when building synthetic data or  mixing existing natural datasets.

In this work, we focus on measuring \textit{acoustic diversity} of a set of speech utterances, i.e.\ measuring the diversity in the distribution of voices, genders, accents, emotions, and background noise. Throughout the paper, we use the term ``facets'' to describe these different aspects of acoustic diversity.

We propose a set of metrics, which we collectively refer to as MAD Speech (\textbf{M}easures of \textbf{A}coustic \textbf{D}iversity of {Speech}). We build these metrics by combining a pretrained speech representation and an aggregation function that is used to calculate the diversity of a sample in the representation space. We experiment with a set of off-the-shelf continuous speech representations such as HuBERT~\cite{hubert}, Wav2Vec-BERT~\cite{w2vb}, and SoundStream~\cite{soundstream} as well as with SpeechSim, a  lightweight representation model related to TRILL~\cite{shor2020towards} and COLA~\cite{contrastive_learning_audio_rep}. We explore the average pairwise cosine dissimilarity and Vendi Score~\cite{vendi} as aggregation functions. 

Further, we learn per-facet projection models with the goal of highlighting contributions of each facet {``independently''} from others. Such a setup is beneficial since it (a) provides a better insight into the relative contributions of different aspects of diversity, and (b) allows to avoid fusing of the different facets in the same metric with uncontrolled relative importance. 

To evaluate the proposed metrics, we build a collection of datasets with known ground-truth acoustic diversity levels. We achieve that by sub-sampling standard speech datasets, while controlling for some measurable parameter that reflects the diversity of the samples along a particular facet (e.g., number of unique voices). Repeating that for all facets, we build a suite of datasets that can be used to measure to what extent a particular metric is capable of measuring acoustic diversity.

\looseness-1Finally, we run a set of studies to measure the impact of various improvements proposed in the literature on the acoustic diversity: fast non-autoregressive decoding of SoundStorm replacing the auto-regressive semantic-to-acoustic model of  AudioLM~\cite{Borsos2023,soundstorm} and Best-of-K decoding of SPEAR-TTS to increase quality of generated audio~\cite{spear-tts}. We also investigate how changes in the temperature sampling affect the acoustic diversity and compare several off-the-shelf TTS systems. These studies highlight that acoustic diversity changes non-trivially across the considered scenarios and it must be taken into account when selecting a model and a sampling setup.


\section{Background}
\paragraph{Diversity in generative models}
Different diversity metrics have been proposed for evaluating generative models in computer vision \citep{FID, pmlr-v119-naeem20a, NEURIPS2018_f7696a9b, NEURIPS2019_0234c510, jalali2023rke} and natural language processing \citep{Zhang*2020BERTScore:, li-etal-2016-diversity, selfbleu, pairwise_bleu, ngram_entropy, shu-etal-2019-generating}. However these metrics are either computed on raw inputs for text \citep{selfbleu, pairwise_bleu, alihosseini-etal-2019-jointly} or based on underlying representations specific to text and images~\citep{FID, cifka2018eval, tevet-berant-2021-evaluating} which makes them unsuitable for evaluating acoustic diversity in generative speech models. 

\paragraph{Diversity metrics for generative audio}
Already the first paper proposing a generative spoken language model (GSLM) recognized the importance of measuring speech diversity~\citep{Lakhotia2021}. However, since GSLM is not able to generate acoustically diverse content (e.g., it supports only a predefined set of speakers), the focus of \citet{Lakhotia2021} was on measuring the diversity on the transcript level. In a follow-up paper, \citet{Kharitonov2022} proposed a modification of GSLM that has a more expressive prosody yet their focus was on improving and measuring within-utterance variability of prosody. To the best of our knowledge, our work is the first to study across-utterance acoustic diversity in speech generation.

\paragraph{Speech representations}
Learning compact and powerful speech representations is an extremely prolific research area~\cite[e.g.,][]{cpc,wav2vec,wav2vec2,contrastive_learning_audio_rep,hubert,kharitonov_cpc_time,Chung2021,data2vec,chen2022wavlm}. Often, generative audio systems are built on top of such representations by appropriately discretizing them~\cite{Lakhotia2021,polyak21_interspeech,audiogen, Kharitonov2022,Borsos2023,audiopalm,defossez2023high}. This informs our choice of representations to experiment with: HuBERT~\citep{hubert} (used in the GSLM family of models~\citep{Lakhotia2021,Polyak2021,Kreuk2021,Kharitonov2022,nguyen2023expresso}) and Wav2Vec-BERT~\citep{Chung2021} (used by the AudioLM lineage~\citep{Borsos2023,spear-tts,soundstorm,musiclm}). We complement these off-the-shelf representations by SpeechSim, a variant of TRILL~\citep{shor2020towards} and COLA~\citep{contrastive_learning_audio_rep} that learns to contrast pairs of speech chunks. In turn, we also learn projection models that map SpeechSim representations into embeddings that are specialized for each specific facet of diversity.

\section{Learning Representations}
\label{sec:learning_representations}
We take a two-steps approach for measuring diversity of a set of speech utterances. First, we encode raw waveform audio in a compact representation and, afterwards, we estimate the diversity of a set of utterances in this embedding space. Hence, it is vital that the information about the relevant properties of speech is preserved in the embedding space, while the irrelevant details are omitted. We calculate embeddings in a hierarchical way: at first, we map audio to generic embeddings and then apply lightweight specialized projection models that are tailored to capture a specific facet. As a result, our approach allows us to get reliable per-facet diversity metrics. More details can be found in Section~\ref{ss:training-details}.

\textbf{SpeechSim} We aim to learn speech representations where acoustically similar audio segments are close to each other in the embedding space. To do so, we train a self-supervised contrastive model with positive pairs coming from non-overlapping fixed-length chunks of the same speech utterance. The main idea being that such speech segments are likely to be acoustically similar. We refer to this model as \textit{SpeechSim}. SpeechSim provides a speech representation that are sensitive to changes in acoustic diversity, however, it has all types of acoustic properties mixed up together. Indeed, our experiments in Section~\ref{ss:experiments} show that the vanilla SpeechSim embeddings are strongly biased towards speaker voice. We therefore learn  projection models (heads) on top of SpeechSim embeddings to extract different types of acoustic information.

\textbf{Learning Projections} We want to extract the signal relevant to each individual facet from our pretrained embedding while suppressing information relevant to other facets. Thus, we train lightweight projection models on top of pretrained SpeechSim embeddings for each facet. To this end, we assume that we have labelled data and train these models with a contrastive objective where positive pairs are sampled from the same class w.r.t the facet.

\section{From Speech Representations to Diversity Metrics}
Once a set of speech utterances is encoded with one of the representations (Section~\ref{sec:learning_representations}), we can measure the diversity within the set in a variety of ways. We focus on two measures: mean pairwise dissimilarity and Vendi Score~\citep{vendi}.

\paragraph{Mean pairwise dissimilarity} 

Denoting the sample size as $n$ and embeddings as $\ve$, we calculate:
\begin{equation}
    1 -\frac{1}{n(n-1)} \sum_{i \neq j; i,j \le n} \frac{\ve_i^T\ve_j}{||\ve_i||\cdot ||\ve_j||}
    \label{eq:cosine}
\end{equation}

We subtract the mean cosine similarity from 1, as it is more convenient to measure diversity (dissimilarity) rather than similarity. This way, an agreement with the ground-truth diversity would result in a positive correlation (see Section~\ref{ss:evaluating-diversity-metric}).

\paragraph{Vendi Score} \citet{vendi} proposed to measure diversity of a sample as the exponential of the Shannon entropy of the eigenvalues of the pairwise similarity matrix, normalized by the number of vectors. In practice, we built it by computing pairwise cosine similarity and dividing by the number of examples. Formally, denoting the eigenvalues of the normalized similarity matrix as $\{\lambda_k\}_{k=1}^n$, Vendi Score is calculated as:
\begin{equation}
\label{eq:vendi}
\exp \left[- \sum_{i=1}^n \lambda_i \log \lambda_i \right]
\end{equation}

\noindent Eq.~\ref{eq:vendi} is well-defined, as the normalized similarity matrix is positive semi-definite and $0 \le \lambda_i \le 1$.

\section{Evaluating Diversity Metrics}
\label{ss:evaluating-diversity-metric}

\subsection{Our Approach}
We break down the acoustic diversity into the following facets: speaker voice, gender, emotion (or style), accent, and background noise. For each of these facets, we repeatedly sample subsets from a large dataset of natural speech utterances while systematically increasing the level of diversity in a controllable way, as further described below. Next, we calculate Spearman's rank correlation \citep{spearman04} between diversity metric scores and the ground-truth diversity level's rank. Metrics with a higher rank correlation are better at representing a particular facet of the acoustic diversity.

In the following, we describe the general template for building a series of datasets of controllable diversity, using voice diversity as a reference. We fix the sample size \textit{a priori} and define voice diversity as the number of distinct speakers in a set of examples. To build one dataset, we sample examples from a natural dataset such that we obtain the required number of speakers and have every speaker represented by the same number of utterances. We create a series of datasets by varying the number of distinct speakers. We also ensure that other facets do not differ from one dataset to another as much as possible so that we make sure we indeed test speaker voice diversity.

This process is similar for other facets. We determine gender diversity as a proportion of female voices; background noise diversity is characterized by the number of different background noise classes; and emotion (accent) diversity is determined by computing the entropy of the distribution of the different emotion (accent) classes.

\subsection{Evaluation data}

For each facet and each diversity level, we randomly sample 100 sets using different seeds.

\paragraph{Speakers voices} We build evaluation (validation) sets from LibriTTS \citep{zen2019libritts} test-clean \& test-other (dev-clean \& dev-other). Each set contains 200 utterances of a single gender (either all male voices or all female voices) to test whether a particular metric is sensitive to the voice diversity \textit{independently} from the gender diversity. We vary the number of distinct voices in $\{5, 10, 15, 20, 25, 33\}$.

To test the influence of gender diversity for the metrics that target voice diversity, we generate additional data where the voice and the gender diversity are pushed in the opposite directions: having a low number of speakers with uniform gender distribution, and a single gender (selected randomly) when the number of speakers is high.

\paragraph{Gender} We build evaluation (validation) sets from LibriTTS \citep{zen2019libritts} test-clean \& test-other (dev-clean \& dev-other). Each set contains 100 utterances. We change the ratio of female voices from 0.0 to 1.0 in increments of 0.1.

We built two types of evaluation samples, one with equal number of different speaker voices in the samples and another one where number of different speaker voices is maximum (resp. minimum) when there is few (resp. lots of) gender variation to test if the gender metric is speaker-sensitive. 
    
\looseness-1\paragraph{Emotion} We build evaluation (validation) sets from 40\% (10\%) of EmoV \citep{adigwe2018emotional} randomly sampled and test (dev) splits of Expresso \citep{nguyen2023expresso}. 
We built two types of samples: (1) with equal number of different speaker voices and (2) with a single speaker voice for datasets with high entropy but all speaker voices---same number of utterances per speaker---for datasets with low entropy. Speakers in test and dev sets are identical to those found in the training set. Emotion classes match in train, dev, and test.
    
\paragraph{Accent} We build evaluation (validation) sets from test (dev) sets of VCTK \citep{vctk}. Accent classes are common between train, dev, and tests sets. Again, we built two types of samples (1) with equal number of different speaker voices and (2) with the minimum number of different speaker voices for high-entropy datasets but all speaker voices--- equally represented---for low-entropy datasets. Speakers in test and dev sets are distinct from those found in the training set. 
    
\paragraph{Background Noise} We build evaluation and validation sets by sampling 20\% and 10\% of AudioSet, respectively~\citep{Gemmeke2017AudioSA}. 
We select examples that have exactly two tags, one of them being ``Speech''. We treat the second tag as a class of the background noise. Next, we construct a sample by randomly selecting 100 examples with varying number of distinct noise classes (1, 5, 10, 25, 50, or 100).

\begin{table*}[tb]
\centering
\resizebox{0.8\textwidth}{!}{
\begin{tabular}{lcccccccc}
\toprule
  &     \multicolumn{2}{c}{Male}                  &        \multicolumn{2}{c}{Female} \\ 
&  Avg. Cosine &      Vendi score &  Avg. Cosine &    Vendi score &  \\ 
\midrule
  
SoundStream  & \ \hphantom{-}0.544 (± 0.312) & \ \hphantom{-}0.759 (± 0.175) & \ \hphantom{-}0.234 (± 0.426) &  \ \hphantom{-}0.731 (± 0.284)  \\

Wav2vec-BERT   & \ -0.448 (± 0.194) & \ -0.419 (± 0.193) & \ -0.118 (± 0.258) & \ -0.094 (± 0.245) \\

HuBERT   & \ \ -0.575 (± 0.094) & \ -0.447 (± 0.202) & \ -0.178 (± 0.291)  &  \ \hphantom{-}0.007 (± 0.321) \\

SpeechSim & \ \hphantom{-}0.811 (± 0.074) & \ \hphantom{-}\textbf{1.000} (± 0.000) & \ \hphantom{-}0.834 (± 0.073)  & \ \hphantom{-}\textbf{1.000} (± 0.000)  \\ 

\midrule
Trained from scratch & \ \hphantom{-}0.480 (± 0.200) & \ \hphantom{-}0.962 (± 0.044) & \ \hphantom{-}0.666 (± 0.200) & \ \hphantom{-}\underline{0.963} (± 0.053)  \\

SpeechSim/Voice & \ \hphantom{-}0.823 (± 0.063) & \ \hphantom{-}\underline{0.993} (± 0.019) & \ \ \hphantom{-}0.881 (± 0.077)  & \ \hphantom{-}\textbf{1.000} (± 0.000) \\
\bottomrule
\end{tabular}
}
\caption{Voice diversity. Average Spearman correlations ($\pm$ standard error) between number of distinct speakers and diversity scores induced by speech representations. Male (Female) refer to samples with male-only (female-only) voices. Best results are in bold, second best results are underlined.}
\label{tab:speakers-tab}
\end{table*}

\begin{table*}
\centering
\resizebox{0.8\textwidth}{!}{
\begin{tabular}{lcccccccc}

\toprule
Female voices ratio  &     \multicolumn{2}{c}{$\le 0.5$}                  &        \multicolumn{2}{c}{$> 0.5$} \\ 
&  Avg. Cosine &      Vendi score &           Avg. Cosine &    Vendi score &  \\ 
\midrule
  
SoundStream &    \ \hphantom{-}0.095 (± 0.463) & \ \hphantom{-}0.092 (± 0.435) & \  \hphantom{-}0.117 (± 0.410)  & \  \hphantom{-}0.097 (± 0.367)  \\ 


Wav2vec-BERT & \ -0.035 (± 0.434)  & \ \hphantom{-}0.026 (± 0.427) &  \ \hphantom{-}0.159 (± 0.406)   & \ \hphantom{-}0.105 (± 0.415)    \\ 
                                    
                                    
HuBERT &  \ \hphantom{-}0.117 (± 0.431)  & \ \hphantom{-}0.223 (± 0.448) &  \ \hphantom{-}0.109 (± 0.445)  & \ \hphantom{-}0.004 (± 0.444)  \\ 
                                    
SpeechSim &  \ \hphantom{-}0.186 (± 0.473)  & \ \hphantom{-}0.132 (± 0.440) & \  \hphantom{-}0.129 (± 0.438)  &  \ \hphantom{-}0.162 (± 0.455)  \\ 

\midrule
Trained from scratch &  \ \hphantom{-}\underline{0.861} (± 0.076)  &  \ \hphantom{-}0.478 (± 0.320) &  \ \hphantom{-}\underline{0.912} (± 0.047)  &  \ \hphantom{-}0.538 (± 0.317)    \\ 

SpeechSim/Gender &  \ \hphantom{-}\textbf{0.902} (± 0.108) & \ \hphantom{-}0.708 (± 0.249) &   \ \hphantom{-}\textbf{0.946} (± 0.057)   & \ \hphantom{-}0.742 (± 0.216) \\
\midrule[\heavyrulewidth] \\
\end{tabular}
}
\caption{Gender diversity. Average Spearman correlations between proportion of female voices and diversity scores induced by speech representations. 
}
\label{tab:gender-tab}
\end{table*}

\section{Experimental Setup}

Details about the training data and the baseline representations we use can be found in Appendix~\ref{ss:data} and Appendix~\ref{ss:baseline_representations} respectively.

\subsection{Training details} \label{ss:training-details}
\paragraph{SpeechSim}
The input to SpeechSim is a 6-seconds chunk of audio with sampling rate of 16kHz. We compute the mel-spectrogram using a window length of 512, hop length of 256 and 156 bins. This produces 376 temporal frames of dimension 156. The model is based on a small ViT architecture~\citep{vit}, containing 12 layers with each layer having 6 attention heads, embedding dimension of 512, and a feed-forward layer of dimension 1024 (25M parameters). The final embeddings are averaged across the time axis and are projected into a 192-dimensional vector.
We train SpeechSim with the semi-hard triplet loss \citep{tripletloss}, using a total batch size of 3840. We train the model for a total of $10^5$ steps.

\looseness-1\paragraph{Projection models}
We train projection models on top of SpeechSim embeddings with. a standard contrastive loss \citep{radford2021learning}. The positive pairs are constructed by taking two examples with the same labels (e.g., same emotion). Except for the gender projector, all models have 2 fully connected layer (of size 256 and 128) with a GELU activation~\citep{hendrycks2016gaussian} in-between. In the case of the gender projector, we found that a 4-layers network works better (dimensions of 256, 256, 128 and 128). Dropout rate of 0.1 is applied to SpeechSim input representations. We trained the projection models using Adam~\citep{kingma2014adam} and use a learning rate of 1e-4, and a batch size of 128. We also apply a weight decay of 1e-3 for projection models related to speaker voices, emotion and background noise and 1e-4 for other facets. We select hyperparameters using validation loss.

All embeddings that we use in our experiments have a single vector representing the entire utterance. To achieve that, we average embeddings over the time axis. We found this to perform better than max-pooling.

\begin{table*}[ht]
\centering
\resizebox{0.8\textwidth}{!}{\begin{tabular}{lccccc}
\toprule
                                                   &            \multicolumn{2}{c}{ EmoV} &    \multicolumn{2}{c}{ Expresso}     \\  
                                                   &   Avg. Cosine &       Vendi score &     Avg. Cosine &      Vendi score \\ \midrule
                                                   
 SoundStream &  \hphantom{-}0.137 (± 0.573) &  \hphantom{-}0.714 (± 0.250) &  \hphantom{-}0.357 (± 0.383) &  \hphantom{-}0.495 (± 0.348)  \\

 Wav2vec-BERT &  \hphantom{-}0.682 (± 0.182) &  \hphantom{-}0.594 (± 0.080) &     \hphantom{-}0.668 (± 0.170) &  \hphantom{-}0.705 (± 0.070)  \\

 HuBERT  &  \hphantom{-}0.785 (± 0.148) &  \hphantom{-}0.683 (± 0.088) &  \hphantom{-}0.725 (± 0.154) &  \hphantom{-}0.725 (± 0.064)   \\ 

 SpeechSim &  \hphantom{-}0.809 (± 0.200) &  \hphantom{-}0.823 (± 0.193)  &  \hphantom{-}0.739 (± 0.249) &  \hphantom{-}0.864 (± 0.128)    \\ 

\midrule
 Trained from scratch &  \hphantom{-}0.991 (± 0.020) &  \hphantom{-}0.962 (± 0.051)   &  \hphantom{-}\underline{0.998} (± 0.008) &  \hphantom{-}0.988 (± 0.018) &   \\ 
 
 Trained from scratch/Expresso &  \hphantom{-}0.498 (± 0.250)  & \hphantom{-}0.513 (± 0.274) & - & -   \\
 
 Trained from scratch/EmoV & - & - & \hphantom{-}0.247 (± 0.562)  &  \hphantom{-}0.251 (± 0.512)  \\

 SpeechSim/Emotion &  \hphantom{-}\textbf{0.995} (± 0.016) &    \hphantom{-}\underline{0.993} (± 0.019)   &  \hphantom{-}\textbf{0.999} (± 0.007) &  \hphantom{-}0.987 (± 0.017) \\ 
 
 \midrule
 
 SpeechSim/Emotion-Expresso  & \hphantom{-}0.641 (± 0.279)  &  \hphantom{-}0.709 (± 0.243) & - & -  \\
 
 SpeechSim/Emotion-EmoV & - & - & \hphantom{-}0.273 (± 0.465) & \hphantom{-}0.321 (± 0.464) \\
 
\bottomrule[\heavyrulewidth]
\end{tabular}
}
\caption{Emotion diversity. Average Spearman correlations between the classes entropy in EmoV and Expresso and diversity scores induced by the speech representations. SpeechSim/Emotion-Expresso (resp. Speech/Emotion-EmoV) refers to SpeechSim emotion head trained on Expresso (resp. EmoV). 
}
\label{tab:emotion-tab}
\end{table*}

\section{Evaluating MAD Speech}
\label{ss:experiments}
In this Section, we use the collection of per-facet diversity benchmarks (Section~\ref{ss:evaluating-diversity-metric}) to assess which speech representations are more suitable for estimating diversity of a set of speech utterances when only one facet is changed (Section~\ref{ss:sensitivity}). Next, we run a series of experiments with two facets changed in the opposite directions (Section~\ref{ss:entangled}). 

\subsection{Variability in a single facet}

\label{ss:sensitivity}

\paragraph{Voice Diversity}

From Table~\ref{tab:speakers-tab} we firstly notice that, across all columns, SpeechSim shows higher correlation scores than off-the-shelf embedding models and the models trained from scratch. Further voice specialization of SpeechSim (SpeechSim/Voice) results in similar scores. Finally, on comparing average cosine vs.\ Vendi Score aggregation, we observe that often Vendi Score shows higher correlations.

\paragraph{Gender diversity} In Table~\ref{tab:gender-tab} we report correlation scores for the gender diversity. Again, we split in two groups: male-voice and female voice-dominant. From the results, we notice that the correlations showed by the non-specialized embeddings are extremely weak. However, both specialized models reach very high correlations (e.g., up to 0.946 for SpeechSim/Gender). All in all, gender-specific projector gets the higher correlation score. In this setup, Vendi Score underperforms w.r.t.\ the average cosine dissimilarity.

\paragraph{Emotion diversity} In Table~\ref{tab:emotion-tab} we report average Spearman correlations for the tested representation models, when tested on EmoV and Expresso separately. From these results, we see that, generally, SpeechSim performs better than other generic representations across all configurations. Equally, the specialized projection of SpeechSim for the emotion facet, SpeechSim/Emotion performs best across all representations. As with voice diversity, Vendi Score gets higher correlations than average cosine dissimilarity.

To make sure that our models can measure diversity of emotions beyond that is covered by the (limited) label sets of EmoV and Expresso, we run an experiment where the projection model is trained on one dataset and tested on another (the only common label is ``neutral'').
Table~\ref{tab:emotion-tab} shows that SpeechSim finetuned on Expresso does transfer to EmoV classes as it obtains a high Spearman correlation. However, SpeechSim finetuned on EmoV gets a Spearman correlation equal to 0.321.
Digging deeper, we additionally compute the correlation of SpeechSim/Emotion-EmoV scores where we average all scores (obtained with different seeds) so that we obtain a single score for each diversity level.
In this, SpeechSim/Emotion-Emov gets a correlation of 0.964 which means that SpeechSim finetuned on EmoV generalizes well on average.

\paragraph{Accent diversity} We report average correlations in Table~\ref{tab:accent-tab}. Here, again, we notice that SpeechSim has higher correlations than other non-specialized embedding models. Both specialized model perform similarly good with perfect correlations. As before, Vendi Score leads to better correlations across all setups.

\begin{table}[tb]
\centering
\resizebox{\columnwidth}{!}{\begin{tabular}{lccccc}
\toprule
                                                   &  Avg. Cosine &     Vendi score  \\ 
\midrule
                                                   
SoundStream & 0.234 (± 0.517) & 0.217 (± 0.581)   \\
Wav2vec-BERT  & 0.425 (± 0.415) & 0.557 (± 0.376)   \\
HuBERT  & 0.372 (± 0.425) & 0.526 (± 0.366)  \\
SpeechSim & 0.446 (± 0.476) & 0.574 (± 0.328)   \\
\midrule
Trained from scratch & \underline{0.995} (± 0.016) & \textbf{1.000} (± 0.000)  \\
SpeechSim/Accent & 0.991 (± 0.021) & \textbf{1.000} (± 0.000)   \\
\bottomrule
\end{tabular}}
\caption{Accent diversity. Average Spearman correlations between the accent class entropy and diversity scores induced by speech representations.}
\label{tab:accent-tab}
\end{table}

\paragraph{Background noise diversity}
We report our results in Table~\ref{tab:background-noise-tab}. We see that generally all representations have somewhat low levels of correlation and high variability across seeds probably due to the fact that AudioSet is pretty noisy and multiple labels can be found in a single example. That said, SpeechSim/Noise combined with Vendi Score obtains the second highest correlation, following HuBERT.

We additionally compute the correlation of the SpeechSim/Noise scores where we average all scores (obtained with different seeds) so that we obtain a single score for each diversity level. In this setup, SpeechSim/Noise gets a Spearman correlation of 1.00, indicating that it works very well on average.

\vspace{4mm}

All in all, Tables~\ref{tab:speakers-tab} to \ref{tab:background-noise-tab} show that SpeechSim generic representations achieve higher correlation scores than off-the-shelf embedding models while having lower latency, this motivates us to rely on it as a basis for per-facet models.

\begin{table}
\centering
\resizebox{\columnwidth}{!}{
\begin{tabular}{lcccccc}
\toprule

&    Avg. Cosine &       Vendi score \\ 
\midrule
SoundStream &   \hphantom{-}0.141 (± 0.390) &   \hphantom{-}0.265 (± 0.433)  \\
                                      
Wav2vec-BERT &  -0.014 (± 0.417) &  -0.070 (± 0.452)     \\

HuBERT &  \hphantom{-}0.218 (± 0.356) &  \hphantom{-}\textbf{0.531} (± 0.332)  \\

SpeechSim &  \hphantom{-}0.254 (± 0.309) &  \hphantom{-}0.466 (± 0.322)  \\

\midrule
                                   
Trained from scratch &  \hphantom{-}0.173 (± 0.355) &   \hphantom{-}0.355 (± 0.404)  \\
SpeechSim/Noise &  \hphantom{-}0.116 (± 0.359) &  \hphantom{-}\underline{0.469} (± 0.413) \\ 

\bottomrule 


                                      
                                    

\end{tabular}}
\caption{Background noise diversity. Average Spearman correlations between the number of different classes of noise and diversity scores induced by speech representations. 
}
\label{tab:background-noise-tab}
\end{table}

\subsection{Are facets mixed within metrics?}
\label{ss:entangled}

Above we assessed how good the proposed scores are in detecting changes in a single factor of diversity with the rest fixed. However, one can expect that in a real-life scenario many factors can change simultaneously. In this Section, we verify whether this behavior takes place.

\looseness-1\paragraph{Gender and Voices} We design a sequence of samples such that the voice and gender diversity are changing in the opposite directions. We continuously change the proportion of female voices from 0.0 to 1.0: the gender diversity grows at first, peaks when the ratio is 0.5, and then starts to decrease. While the gender diversity increases, we reduce the number of speakers (from 30 to 2) and when the gender diversity starts to fall, we increase the number of speakers back (from 2 to 30).

We report results in Table~\ref{tab:gender_control_speakers_tab}. From this results we see that SpeechSim/Gender is positively correlated with the gender diversity. In contrast, the SpeechSim/Voice metric tracks the speaker diversity and hence is negatively correlated with the gender diversity. Finally, the original SpeechSim embeddings have negative correlation with the gender diversity. Moreover, its correlation coefficient is smaller than that of SpeechSim/Voice. From that we conclude that the vanilla SpeechSim measure overemphasizes voice over gender diversity.

\begin{table}[tb]
\centering
\resizebox{\linewidth}{!}{\begin{tabular}{lcc}
\toprule
&    Avg. cosine               &       Vendi score  \\ \midrule

\multicolumn{2}{c}{$< 0.5$ female voices} \\
\midrule
Speech Sim & -0.709 (± 0.108)  &  -0.734 (± 0.107)     \\


SpeechSim/Speaker & -0.657 (± 0.088)  &  -0.686 (± 0.092)  \\ 
SpeechSim/Gender &  \hphantom{-}{0.781} (± 0.259)  &  \hphantom{-}0.147 (± 0.341)    \\

\midrule
 \multicolumn{2}{c}{$\ge 0.5$ female voices} \\
 \midrule
SpeechSim & -0.706 (± 0.110) &  -0.709 (± 0.120)     \\
SpeechSim/Speaker & -0.648 (± 0.080) & -0.668 (± 0.092)  \\
SpeechSim/Gender &   \hphantom{-}{0.827} (± 0.285) &  \hphantom{-}0.313 (± 0.342)  \\
\bottomrule
\end{tabular}}
\caption{Changing voice and gender diversity in the opposite directions. Average Spearman correlation between the gender diversity and the diversity scores induced by speech representations.} 
\label{tab:gender_control_speakers_tab}
\end{table}

\begin{figure}[tb]
 \centering\small
\subcaptionbox{SpeechSim/Gender, avg.\ pairwise dissim. 
\label{fig:madspeech-gender-gender-control}}{\includegraphics[width=0.47\linewidth]{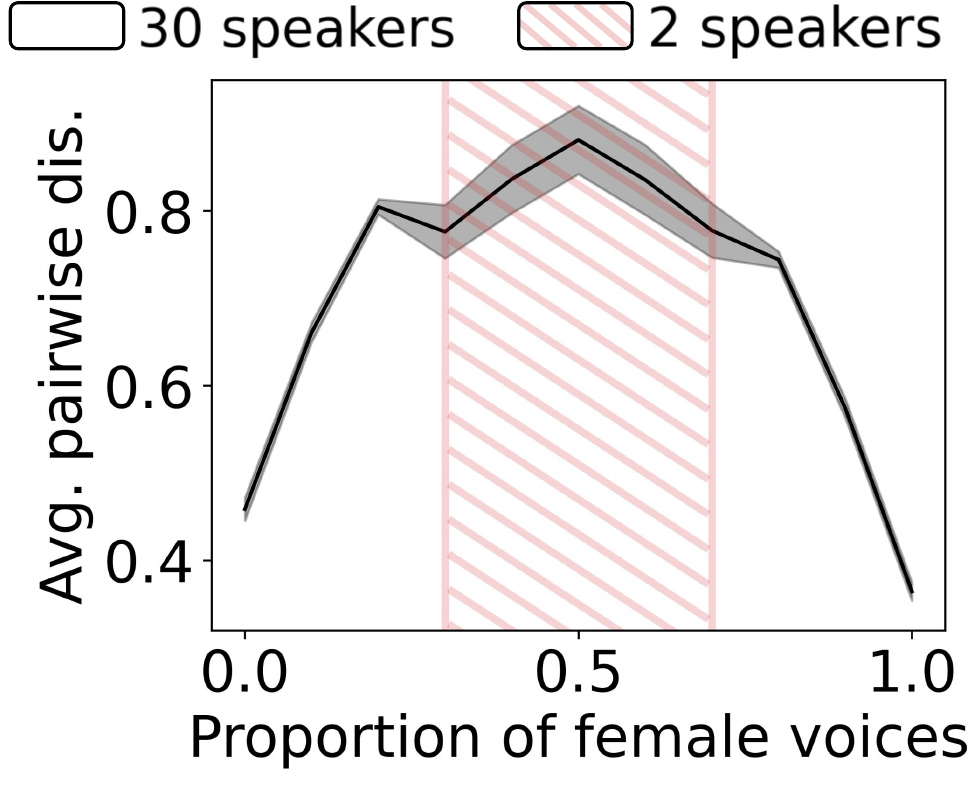}}%
\hspace{.7em}
\subcaptionbox{SpeechSim/Voice, Vendi score. \label{fig:madspeech-speakers-gender-control}}{\includegraphics[width=0.47\linewidth]{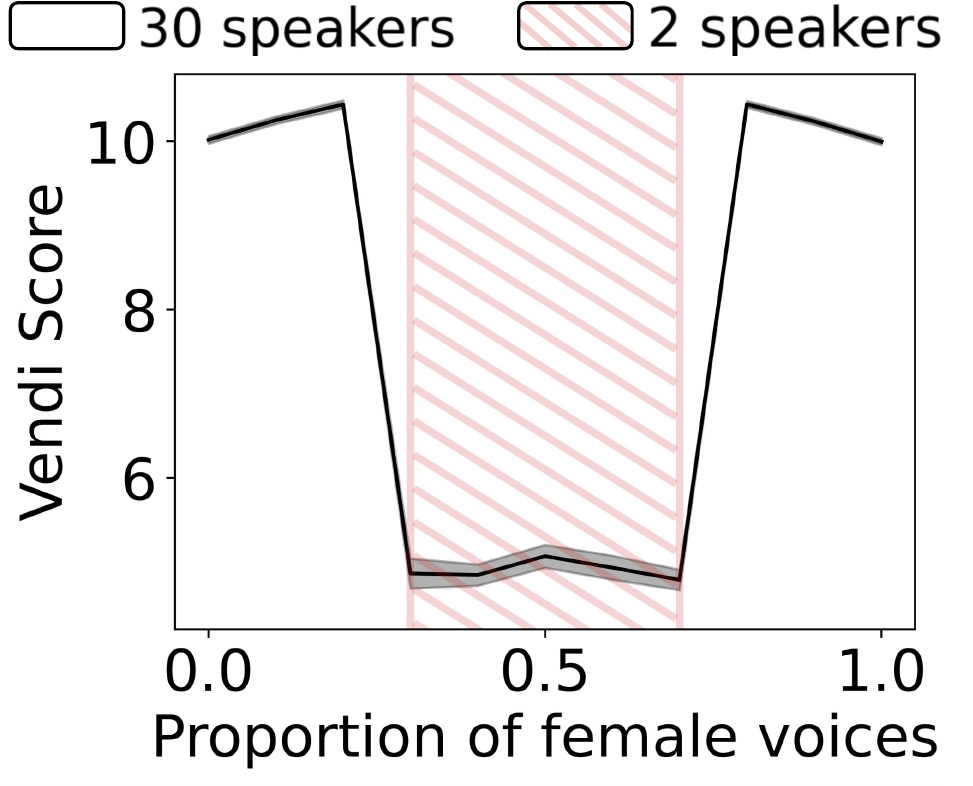}}
\caption{Changing the proportion of female voices and number of distinct speaker voices simultaneously. Grey area represents 95\% CI.}
\label{fig:gender-control-speakers}
\end{figure}

We further illustrate this experiment in Figure~\ref{fig:gender-control-speakers}, reporting SpeechSim/Gender and SpeechSim/Voice scores as the female voice ratio changes. From Figure~\ref{fig:gender-control-speakers} we see that SpeechSim/Gender is almost invariant to the changes in the number of speakers. Conversely, SpeechSim/Voice is relatively insensitive to changes of gender diversity.

We run a similar experiment for accents and emotions and report it in Appendix~\ref{ss:facets-mix}. We observe similar findings, raw SpeechSim is highly influenced by Speaker voice variations while specialized facets are unaffected by changes in voices.

\vspace{2mm}
Overall, the experiments reported in this Section and Appendix~\ref{ss:facets-mix} support the necessity of introducing per-facet projection models. While SpeechSim embeddings might provide reasonably good correlations with the ground-truth diversity when only one facet is changed, they implicitly mix all facets into a single score.
We find this inconvenient, as (a) changes in one facet can mask changes in another, (b) the user might want to control the trade-off explicitly.
Hence, per-facet models are preferrable.

\section{Comparing Generative Models}
In this section, we show that despite that the acoustic diversity was mostly ignored in the literature when proposing new models and approaches, it does change.

We start by assessing whether the introduction of SoundStorm~\citep{soundstorm} to the AudioLM framework~\cite{Borsos2023} lead to any hidden penalties in acoustic diversity (Section~\ref{ss:audiolm-vs-soundstorm}) before comparing acoustic diversity of off-the-shelf TTS models (Section~\ref{ss:tts}). Additional studies on the effect of the Best-K sampling technique used by~\citet{spear-tts} on acoustic diversity and how temperature changes acoustic diversity of the synthesized speech can be found in Appendix~\ref{ss:bestk} and Appendix~\ref{ss:temp}. In this Section, we use the same terminology as \citet{Borsos2023} where semantic tokens refer to tokens learned with a self-supervised masked language modeling objective \citep{w2vb, hubert} while acoustic tokens refer to tokens learned with a full reconstruction objective from a residual vector quantizer \citep{soundstream}.

\subsection{Semantic-to-Acoustic Token Conversion: SoundStorm vs.\ AudioLM}
\label{ss:audiolm-vs-soundstorm}
Generating acoustic tokens from semantic tokens is a core part of AudioLM~\cite{Borsos2023}. This step is believed to be responsible for modelling the acoustic details of the generated audio, such as voice, background noise, etc. \citet{soundstorm} proposed to implement it with a non-autoregressive Transformer, yielding the SoundStorm model that dramatically improved inference time. However, it is unclear whether this change harms acoustic diversity of the generated audio.

We take 32 male- and 32 female-voice utterances from hold-out LibriTTS test-clean and represent them as sequences of semantic tokens. Next, we convert these semantic token sequences to acoustic token sequences using both models, repeating the process 128 times. We convert the acoustic tokens into audio using the corresponding SoundStream codec~\cite{soundstream}. Finally, for each combination of a source utterance and a model, we calculate the diversity of the produced  audio samples.  We use exact same SoundStorm and AudioLM models as~\citet{soundstorm}.

In Table~\ref{tab:relative_diversity} we report, for each facet of the acoustic diversity, the ratio of utterances where AudioLM generated higher-diversity audio than SoundStorm. We see that SoundStorm  produces audio that is more diverse in voices, emotions, accents, and background noise, while being behind in the gender diversity. 

\begin{table}
\centering
\resizebox{\linewidth}{!}{
\begin{tabular}{lcccc}
\toprule
Voice & Gender & Emotion & Accent & Background noise \\
\midrule
0.0 & 62.5 & 6.3 & 1.6 & 4.7 \\

\bottomrule
\end{tabular}
}
\caption{Semantic-to-acoustic tokens resynthesis task. The proportion of sets of utterances (\%) where the AudioLM stage 2 model generates samples that are  more diverse than those generated by SoundStorm.}
\label{tab:relative_diversity}
\end{table}

\subsection{Acoustic diversity in Text-to-Speech systems}\label{ss:tts}
Next, we measure acoustic diversity of several publicly available TTS systems. Here, our goal is to show that MAD Speech brings a new dimension to the evaluation of real-world TTS systems. We compare \textit{Bark TTS}\footnote{\url{https://github.com/suno-ai/bark}}, \textit{StyleTTS 2} \citep{styletts2}, \textit{Tortoise TTS} \citep{betker2023better} and \textit{FastSpeech 2} \citep{Ren2020}. Details about these models can be found in Appendix~\ref{ss:tts-models}. As some of these systems are deterministic, we adopt a different setup and synthesize all transcripts in shuffled LibriTTS test-clean and calculate the average acoustic diversity across batches of audio of size 128. We report the results in Table~\ref{tab:tts}.

\begin{table}[t]
\centering
\resizebox{\linewidth}{!}{
\begin{tabular}{lrrrrrr}
\toprule
 & Voice & Gender & Emotion & Accent & Back.\ Noise \\
 \midrule
Bark TTS & 39.41 & 0.90 & 8.29 & 8.30 & 3.84 \\
Tortoise TTS & 30.94 & 0.92 & 8.73 & 8.17 & 3.17\\
StyleTTS 2 & 31.54 & 0.37 & 7.73 & 6.81 & 2.60  \\
FastSpeech 2 & 19.33 & 0.29 & 6.74 & 5.63 & 2.42\\
\bottomrule
\end{tabular}
}
\caption{Acoustic diversity scores for off-the-shelf TTS systems (higher is more diverse). We report Vendi score for all facets except for gender, which uses cosine dissimilarity.}
\label{tab:tts}
\end{table}

We see that Bark and Tortoise TTS are the most diverse across all facets, likely due to training on more diverse internet data. Interestingly, Tortoise TTS tends to be more diverse than Bark in emotions while less or equally diverse in other facets. StyleTTS~2 is considerably less diverse in the gender facet --- just as one would expect, given that the generation is restricted to female voices. FastSpeech 2 is the least acoustically diverse mainly due to being trained on a very homogeneous dataset. 
\vspace{1mm}

All in all, the comparisons we carried out in this Section and Appendix~\ref{ss:effect-diversity} show that previously proposed changes lead to non-trivial and sometimes even non-monotonic changes in per-facet acoustic diversity. Equally, the sampling temperature also affects the resulting acoustic diversity. We also found that MAD Speech can be useful in detecting acoustic diversity trends in off-the-shelf TTS systems. Overall, we believe those findings highlight the necessity for including acoustic diversity metrics in the standard evaluations.

\section{Conclusions \& Future work}
\looseness-1We introduced MAD Speech, a set of metrics for evaluating acoustic diversity in speech. MAD Speech metrics are calculated in two steps: (1) a set of utterances is mapped to an embedding space, (2) acoustic diversity is estimated by comparing the resulting embeddings with an aggregation function (cosine dissimilarity or Vendi Score). Calculating embeddings is done by firstly applying a generic representation model with a subsequent metric-specific projection. We focus on five facets of acoustic diversity: voices, gender, emotion, accents, and background noise.

In order to validate the proposed metrics, we built a collection of datasets with a controlled level of acoustic diversity. Having these sets allows us to evaluate a metric by calculating its Spearman rank correlation with the ground-truth acoustic diversity.

Our empirical study demonstrated that our proposed metrics have a highest agreement with the ground-truth diversity levels when compared to baseline approaches and that they are insensitive to variations of other facets of diversity.

Finally, we highlighted a necessity of controlling acoustic diversity when developing new models and approaches. To this end, we took some recent modelling improvements and demonstrated that they, in fact, affect acoustic diversity of speech. 


\section*{Limitations}\label{s:limitations}
By building MAD Speech, we make it possible to holistically assess progress in generative speech and reveal possible hidden biases in the acoustic scenes. Yet, our metrics are machine-learned models themselves and are limited by the nature of the data we used to build them. Specifically, the public datasets we use are limited to English language. The labels in the data can be noisy and only represent a limited reflection of the world. We also made the choice to treat gender as a binary variable to be consistent with existing datasets, this can lead to minor measurement inaccuracies when facing ambiguously-gendered voices \citep{stoidis22_interspeech}.


\vspace{3mm}
\textbf{Acknowledgements} The authors are grateful to Aren Jansen and Salah Zaiem for their feedback.

\bibliography{custom}

\begin{thebibliography}{69}
\providecommand{\natexlab}[1]{#1}

\bibitem[{Adigwe et~al.(2018)Adigwe, Tits, Haddad, Ostadabbas, and
  Dutoit}]{adigwe2018emotional}
Adaeze Adigwe, No{\'e} Tits, Kevin~El Haddad, Sarah Ostadabbas, and Thierry
  Dutoit. 2018.
\newblock The emotional voices database: Towards controlling the emotion
  dimension in voice generation systems.
\newblock \emph{arXiv preprint arXiv:1806.09514}.

\bibitem[{Agostinelli et~al.(2023)Agostinelli, Denk, Borsos, Engel, Verzetti,
  Caillon, Huang, Jansen, Roberts, Tagliasacchi et~al.}]{musiclm}
Andrea Agostinelli, Timo~I Denk, Zal{\'a}n Borsos, Jesse Engel, Mauro Verzetti,
  Antoine Caillon, Qingqing Huang, Aren Jansen, Adam Roberts, Marco
  Tagliasacchi, et~al. 2023.
\newblock {MusicLM}: Generating music from text.
\newblock \emph{arXiv preprint arXiv:2301.11325}.

\bibitem[{Alihosseini et~al.(2019)Alihosseini, Montahaei, and
  Soleymani~Baghshah}]{alihosseini-etal-2019-jointly}
Danial Alihosseini, Ehsan Montahaei, and Mahdieh Soleymani~Baghshah. 2019.
\newblock \href {https://doi.org/10.18653/v1/W19-2311} {Jointly measuring
  diversity and quality in text generation models}.
\newblock In \emph{Proceedings of the Workshop on Methods for Optimizing and
  Evaluating Neural Language Generation}, pages 90--98, Minneapolis, Minnesota.
  Association for Computational Linguistics.

\bibitem[{Ardila et~al.(2020)Ardila, Branson, Davis, Kohler, Meyer, Henretty,
  Morais, Saunders, Tyers, and Weber}]{common-voice}
Rosana Ardila, Megan Branson, Kelly Davis, Michael Kohler, Josh Meyer, Michael
  Henretty, Reuben Morais, Lindsay Saunders, Francis Tyers, and Gregor Weber.
  2020.
\newblock \href {https://aclanthology.org/2020.lrec-1.520} {Common voice: A
  massively-multilingual speech corpus}.
\newblock In \emph{Proceedings of the Twelfth Language Resources and Evaluation
  Conference}, pages 4218--4222, Marseille, France. European Language Resources
  Association.

\bibitem[{Baevski et~al.(2022)Baevski, Hsu, Xu, Babu, Gu, and Auli}]{data2vec}
Alexei Baevski, Wei-Ning Hsu, Qiantong Xu, Arun Babu, Jiatao Gu, and Michael
  Auli. 2022.
\newblock \href {https://arxiv.org/abs/2202.03555} {data2vec: A general
  framework for self-supervised learning in speech, vision and language}.
\newblock \emph{Preprint}, arXiv:2202.03555.

\bibitem[{Baevski et~al.(2020)Baevski, Zhou, Mohamed, and Auli}]{wav2vec2}
Alexei Baevski, Henry Zhou, Abdelrahman Mohamed, and Michael Auli. 2020.
\newblock \href {https://arxiv.org/abs/2006.11477} {wav2vec 2.0: A framework
  for self-supervised learning of speech representations}.
\newblock \emph{Preprint}, arXiv:2006.11477.

\bibitem[{Bakhturina et~al.(2021)Bakhturina, Lavrukhin, Ginsburg, and
  Zhang}]{bakhturina2021}
Evelina Bakhturina, Vitaly Lavrukhin, Boris Ginsburg, and Yang Zhang. 2021.
\newblock Hi-fi multi-speaker english tts dataset.
\newblock \emph{arXiv preprint arXiv:2104.01497}.

\bibitem[{Betker(2023{\natexlab{a}})}]{betker2023better}
James Betker. 2023{\natexlab{a}}.
\newblock Better speech synthesis through scaling.
\newblock \emph{arXiv preprint arXiv:2305.07243}.

\bibitem[{Betker(2023{\natexlab{b}})}]{tortoisetts}
James Betker. 2023{\natexlab{b}}.
\newblock \href {https://arxiv.org/abs/2305.07243} {Better speech synthesis
  through scaling}.
\newblock \emph{Preprint}, arXiv:2305.07243.

\bibitem[{Borsos et~al.(2023{\natexlab{a}})Borsos, Marinier, Vincent,
  Kharitonov, Pietquin, Sharifi, Roblek, Teboul, Grangier, Tagliasacchi
  et~al.}]{Borsos2023}
Zal{\'a}n Borsos, Rapha{\"e}l Marinier, Damien Vincent, Eugene Kharitonov,
  Olivier Pietquin, Matt Sharifi, Dominik Roblek, Olivier Teboul, David
  Grangier, Marco Tagliasacchi, et~al. 2023{\natexlab{a}}.
\newblock {AudioLM}: a language modeling approach to audio generation.
\newblock \emph{IEEE/ACM Transactions on Audio, Speech, and Language
  Processing}.

\bibitem[{Borsos et~al.(2023{\natexlab{b}})Borsos, Sharifi, Vincent,
  Kharitonov, Zeghidour, and Tagliasacchi}]{soundstorm}
Zal{\'a}n Borsos, Matt Sharifi, Damien Vincent, Eugene Kharitonov, Neil
  Zeghidour, and Marco Tagliasacchi. 2023{\natexlab{b}}.
\newblock Soundstorm: Efficient parallel audio generation.
\newblock \emph{arXiv preprint arXiv:2305.09636}.

\bibitem[{Chen et~al.(2022{\natexlab{a}})Chen, Wang, Chen, Wu, Liu, Chen, Li,
  Kanda, Yoshioka, Xiao, Wu, Zhou, Ren, Qian, Qian, Wu, Zeng, Yu, and
  Wei}]{wavlm}
Sanyuan Chen, Chengyi Wang, Zhengyang Chen, Yu~Wu, Shujie Liu, Zhuo Chen, Jinyu
  Li, Naoyuki Kanda, Takuya Yoshioka, Xiong Xiao, Jian Wu, Long Zhou, Shuo Ren,
  Yanmin Qian, Yao Qian, Jian Wu, Michael Zeng, Xiangzhan Yu, and Furu Wei.
  2022{\natexlab{a}}.
\newblock {WavLM}: Large-scale self-supervised pre-training for full stack
  speech processing.
\newblock \emph{{IEEE} JSTSP}, 16(6).

\bibitem[{Chen et~al.(2022{\natexlab{b}})Chen, Wang, Chen, Wu, Liu, Chen, Li,
  Kanda, Yoshioka, Xiao et~al.}]{chen2022wavlm}
Sanyuan Chen, Chengyi Wang, Zhengyang Chen, Yu~Wu, Shujie Liu, Zhuo Chen, Jinyu
  Li, Naoyuki Kanda, Takuya Yoshioka, Xiong Xiao, et~al. 2022{\natexlab{b}}.
\newblock Wavlm: Large-scale self-supervised pre-training for full stack speech
  processing.
\newblock \emph{IEEE Journal of Selected Topics in Signal Processing},
  16(6):1505--1518.

\bibitem[{Chung et~al.(2021{\natexlab{a}})Chung, Zhang, Han, Chiu, Qin, Pang,
  and Wu}]{w2vb}
Yu-An Chung, Yu~Zhang, Wei Han, Chung-Cheng Chiu, James Qin, Ruoming Pang, and
  Yonghui Wu. 2021{\natexlab{a}}.
\newblock {W2V-BERT}: Combining contrastive learning and masked language
  modeling for self-supervised speech pre-training.
\newblock In \emph{2021 IEEE Automatic Speech Recognition and Understanding
  Workshop (ASRU)}, pages 244--250. IEEE.

\bibitem[{Chung et~al.(2021{\natexlab{b}})Chung, Zhang, Han, Chiu, Qin, Pang,
  and Wu}]{Chung2021}
Yu-An Chung, Yu~Zhang, Wei Han, Chung-Cheng Chiu, James Qin, Ruoming Pang, and
  Yonghui Wu. 2021{\natexlab{b}}.
\newblock {W2V-Bert}: Combining contrastive learning and masked language
  modeling for self-supervised speech pre-training.
\newblock In \emph{ASRU}.

\bibitem[{Cideron et~al.(2024)Cideron, Girgin, Verzetti, Vincent, Kastelic,
  Borsos, McWilliams, Ungureanu, Bachem, Pietquin, Geist, Hussenot, Zeghidour,
  and Agostinelli}]{musicrl}
Geoffrey Cideron, Sertan Girgin, Mauro Verzetti, Damien Vincent, Matej
  Kastelic, Zalán Borsos, Brian McWilliams, Victor Ungureanu, Olivier Bachem,
  Olivier Pietquin, Matthieu Geist, Léonard Hussenot, Neil Zeghidour, and
  Andrea Agostinelli. 2024.
\newblock \href {https://arxiv.org/abs/2402.04229} {Musicrl: Aligning music
  generation to human preferences}.
\newblock \emph{Preprint}, arXiv:2402.04229.

\bibitem[{C{\'\i}fka et~al.(2018)C{\'\i}fka, Severyn, Alfonseca, and
  Filippova}]{cifka2018eval}
Ond{\v{r}}ej C{\'\i}fka, Aliaksei Severyn, Enrique Alfonseca, and Katja
  Filippova. 2018.
\newblock Eval all, trust a few, do wrong to none: Comparing sentence
  generation models.
\newblock \emph{arXiv preprint arXiv:1804.07972}.

\bibitem[{D{\'e}fossez et~al.(2023)D{\'e}fossez, Copet, Synnaeve, and
  Adi}]{defossez2023high}
Alexandre D{\'e}fossez, Jade Copet, Gabriel Synnaeve, and Yossi Adi. 2023.
\newblock \href {https://openreview.net/forum?id=ivCd8z8zR2} {High fidelity
  neural audio compression}.
\newblock \emph{Transactions on Machine Learning Research}.
\newblock Featured Certification, Reproducibility Certification.

\bibitem[{Dosovitskiy et~al.(2020)Dosovitskiy, Beyer, Kolesnikov, Weissenborn,
  Zhai, Unterthiner, Dehghani, Minderer, Heigold, Gelly et~al.}]{vit}
Alexey Dosovitskiy, Lucas Beyer, Alexander Kolesnikov, Dirk Weissenborn,
  Xiaohua Zhai, Thomas Unterthiner, Mostafa Dehghani, Matthias Minderer, Georg
  Heigold, Sylvain Gelly, et~al. 2020.
\newblock An image is worth 16x16 words: Transformers for image recognition at
  scale.
\newblock \emph{arXiv preprint arXiv:2010.11929}.

\bibitem[{Dušek et~al.(2019)Dušek, Novikova, and Rieser}]{ngram_entropy}
Ondřej Dušek, Jekaterina Novikova, and Verena Rieser. 2019.
\newblock \href {https://doi.org/10.1016/j.csl.2019.06.009} {Evaluating the
  state-of-the-art of end-to-end natural language generation: The e2e nlg
  challenge}.
\newblock \emph{Computer Speech \& Language}, 59.

\bibitem[{Friedman and Dieng(2022)}]{vendi}
Dan Friedman and Adji~Bousso Dieng. 2022.
\newblock The vendi score: A diversity evaluation metric for machine learning.
\newblock \emph{arXiv preprint arXiv:2210.02410}.

\bibitem[{Gemmeke et~al.(2017)Gemmeke, Ellis, Freedman, Jansen, Lawrence,
  Moore, Plakal, and Ritter}]{Gemmeke2017AudioSA}
Jort~F. Gemmeke, Daniel P.~W. Ellis, Dylan Freedman, Aren Jansen, Wade
  Lawrence, R.~Channing Moore, Manoj Plakal, and Marvin Ritter. 2017.
\newblock \href {https://api.semanticscholar.org/CorpusID:21519176} {Audio set:
  An ontology and human-labeled dataset for audio events}.
\newblock \emph{2017 IEEE International Conference on Acoustics, Speech and
  Signal Processing (ICASSP)}, pages 776--780.

\bibitem[{Hassid et~al.(2024)Hassid, Remez, Nguyen, Gat, Conneau, Kreuk, Copet,
  Defossez, Synnaeve, Dupoux et~al.}]{Hassid2024}
Michael Hassid, Tal Remez, Tu~Anh Nguyen, Itai Gat, Alexis Conneau, Felix
  Kreuk, Jade Copet, Alexandre Defossez, Gabriel Synnaeve, Emmanuel Dupoux,
  et~al. 2024.
\newblock Textually pretrained speech language models.
\newblock \emph{Advances in Neural Information Processing Systems}, 36.

\bibitem[{Hendrycks and Gimpel(2016)}]{hendrycks2016gaussian}
Dan Hendrycks and Kevin Gimpel. 2016.
\newblock Gaussian error linear units (gelus).
\newblock \emph{arXiv preprint arXiv:1606.08415}.

\bibitem[{Heusel et~al.(2017)Heusel, Ramsauer, Unterthiner, Nessler, and
  Hochreiter}]{FID}
Martin Heusel, Hubert Ramsauer, Thomas Unterthiner, Bernhard Nessler, and Sepp
  Hochreiter. 2017.
\newblock \href
  {https://proceedings.neurips.cc/paper_files/paper/2017/file/8a1d694707eb0fefe65871369074926d-Paper.pdf}
  {Gans trained by a two time-scale update rule converge to a local nash
  equilibrium}.
\newblock In \emph{Advances in Neural Information Processing Systems},
  volume~30. Curran Associates, Inc.

\bibitem[{Hsu et~al.(2021)Hsu, Bolte, Tsai, Lakhotia, Salakhutdinov, and
  Mohamed}]{hubert}
Wei-Ning Hsu, Benjamin Bolte, Yao-Hung~Hubert Tsai, Kushal Lakhotia, Ruslan
  Salakhutdinov, and Abdelrahman Mohamed. 2021.
\newblock Hubert: Self-supervised speech representation learning by masked
  prediction of hidden units.
\newblock \emph{IEEE/ACM Transactions on Audio, Speech, and Language
  Processing}, 29:3451--3460.

\bibitem[{Ito and Johnson(2017)}]{ljspeech}
Keith Ito and Linda Johnson. 2017.
\newblock The {LJ} speech dataset.
\newblock \url{https://keithito.com/LJ-Speech-Dataset/}.

\bibitem[{Jalali et~al.(2023)Jalali, Li, and Farnia}]{jalali2023rke}
Mohammad Jalali, Cheuk~Ting Li, and Farzan Farnia. 2023.
\newblock \href {https://openreview.net/forum?id=PdZhf6PiAb} {An
  information-theoretic evaluation of generative models in learning multi-modal
  distributions}.
\newblock In \emph{Thirty-seventh Conference on Neural Information Processing
  Systems}.

\bibitem[{Kahn et~al.(2020)Kahn, Rivi{\`e}re, Zheng, Kharitonov, Xu,
  Mazar{\'e}, Karadayi, Liptchinsky, Collobert, Fuegen et~al.}]{Kahn2020}
Jacob Kahn, Morgane Rivi{\`e}re, Weiyi Zheng, Evgeny Kharitonov, Qiantong Xu,
  Pierre-Emmanuel Mazar{\'e}, Julien Karadayi, Vitaliy Liptchinsky, Ronan
  Collobert, Christian Fuegen, et~al. 2020.
\newblock {Libri-Light}: A benchmark for {ASR} with limited or no supervision.
\newblock In \emph{IEEE ICASSP}.

\bibitem[{Kharitonov et~al.(2022)Kharitonov, Lee, Polyak, Adi, Copet, Lakhotia,
  Nguyen, Riviere, Mohamed, Dupoux, and Hsu}]{Kharitonov2022}
Eugene Kharitonov, Ann Lee, Adam Polyak, Yossi Adi, Jade Copet, Kushal
  Lakhotia, Tu~Anh Nguyen, Morgane Riviere, Abdelrahman Mohamed, Emmanuel
  Dupoux, and Wei-Ning Hsu. 2022.
\newblock Text-free prosody-aware generative spoken language modeling.
\newblock In \emph{ACL}.

\bibitem[{Kharitonov et~al.(2021)Kharitonov, Rivière, Synnaeve, Wolf, Mazaré,
  Douze, and Dupoux}]{kharitonov_cpc_time}
Eugene Kharitonov, Morgane Rivière, Gabriel Synnaeve, Lior Wolf,
  Pierre-Emmanuel Mazaré, Matthijs Douze, and Emmanuel Dupoux. 2021.
\newblock \href {https://doi.org/10.1109/SLT48900.2021.9383605} {Data
  augmenting contrastive learning of speech representations in the time
  domain}.
\newblock In \emph{2021 IEEE Spoken Language Technology Workshop (SLT)}, pages
  215--222.

\bibitem[{Kharitonov et~al.(2023)Kharitonov, Vincent, Borsos, Marinier, Girgin,
  Pietquin, Sharifi, Tagliasacchi, and Zeghidour}]{spear-tts}
Eugene Kharitonov, Damien Vincent, Zal{\'a}n Borsos, Rapha{\"e}l Marinier,
  Sertan Girgin, Olivier Pietquin, Matt Sharifi, Marco Tagliasacchi, and Neil
  Zeghidour. 2023.
\newblock Speak, read and prompt: High-fidelity text-to-speech with minimal
  supervision.
\newblock \emph{arXiv preprint arXiv:2302.03540}.

\bibitem[{Kingma and Ba(2014)}]{kingma2014adam}
Diederik~P Kingma and Jimmy Ba. 2014.
\newblock Adam: A method for stochastic optimization.
\newblock \emph{arXiv preprint arXiv:1412.6980}.

\bibitem[{Kirk et~al.(2023)Kirk, Mediratta, Nalmpantis, Luketina, Hambro,
  Grefenstette, and Raileanu}]{kirk2023understanding}
Robert Kirk, Ishita Mediratta, Christoforos Nalmpantis, Jelena Luketina, Eric
  Hambro, Edward Grefenstette, and Roberta Raileanu. 2023.
\newblock Understanding the effects of rlhf on llm generalisation and
  diversity.
\newblock \emph{arXiv preprint arXiv:2310.06452}.

\bibitem[{Kreuk et~al.(2021)Kreuk, Polyak, Copet, Kharitonov, Nguyen,
  Rivi{\`e}re, Hsu, Mohamed, Dupoux, and Adi}]{Kreuk2021}
Felix Kreuk, Adam Polyak, Jade Copet, Eugene Kharitonov, Tu-Anh Nguyen, Morgane
  Rivi{\`e}re, Wei-Ning Hsu, Abdelrahman Mohamed, Emmanuel Dupoux, and Yossi
  Adi. 2021.
\newblock Textless speech emotion conversion using decomposed and discrete
  representations.
\newblock \emph{arXiv preprint arXiv:2111.07402}.

\bibitem[{Kreuk et~al.(2022)Kreuk, Synnaeve, Polyak, Singer, D{\'e}fossez,
  Copet, Parikh, Taigman, and Adi}]{audiogen}
Felix Kreuk, Gabriel Synnaeve, Adam Polyak, Uriel Singer, Alexandre
  D{\'e}fossez, Jade Copet, Devi Parikh, Yaniv Taigman, and Yossi Adi. 2022.
\newblock Audiogen: Textually guided audio generation.
\newblock \emph{arXiv preprint arXiv:2209.15352}.

\bibitem[{Kynk\"{a}\"{a}nniemi et~al.(2019)Kynk\"{a}\"{a}nniemi, Karras, Laine,
  Lehtinen, and Aila}]{NEURIPS2019_0234c510}
Tuomas Kynk\"{a}\"{a}nniemi, Tero Karras, Samuli Laine, Jaakko Lehtinen, and
  Timo Aila. 2019.
\newblock \href
  {https://proceedings.neurips.cc/paper_files/paper/2019/file/0234c510bc6d908b28c70ff313743079-Paper.pdf}
  {Improved precision and recall metric for assessing generative models}.
\newblock In \emph{Advances in Neural Information Processing Systems},
  volume~32. Curran Associates, Inc.

\bibitem[{Lakhotia et~al.(2021)Lakhotia, Kharitonov, Hsu, Adi, Polyak, Bolte,
  Nguyen, Copet, Baevski, Mohamed et~al.}]{Lakhotia2021}
Kushal Lakhotia, Eugene Kharitonov, Wei-Ning Hsu, Yossi Adi, Adam Polyak,
  Benjamin Bolte, Tu-Anh Nguyen, Jade Copet, Alexei Baevski, Abdelrahman
  Mohamed, et~al. 2021.
\newblock On generative spoken language modeling from raw audio.
\newblock \emph{TACL}.

\bibitem[{Le et~al.(2023)Le, Vyas, Shi, Karrer, Sari, Moritz, Williamson,
  Manohar, Adi, Mahadeokar, and Hsu}]{voicebox}
Matthew Le, Apoorv Vyas, Bowen Shi, Brian Karrer, Leda Sari, Rashel Moritz,
  Mary Williamson, Vimal Manohar, Yossi Adi, Jay Mahadeokar, and Wei-Ning Hsu.
  2023.
\newblock \href
  {https://proceedings.neurips.cc/paper_files/paper/2023/file/2d8911db9ecedf866015091b28946e15-Paper-Conference.pdf}
  {Voicebox: Text-guided multilingual universal speech generation at scale}.
\newblock In \emph{Advances in Neural Information Processing Systems},
  volume~36, pages 14005--14034. Curran Associates, Inc.

\bibitem[{Li et~al.(2016)Li, Galley, Brockett, Gao, and
  Dolan}]{li-etal-2016-diversity}
Jiwei Li, Michel Galley, Chris Brockett, Jianfeng Gao, and Bill Dolan. 2016.
\newblock \href {https://doi.org/10.18653/v1/N16-1014} {A diversity-promoting
  objective function for neural conversation models}.
\newblock In \emph{Proceedings of the 2016 Conference of the North {A}merican
  Chapter of the Association for Computational Linguistics: Human Language
  Technologies}, pages 110--119, San Diego, California. Association for
  Computational Linguistics.

\bibitem[{Li et~al.(2024)Li, Han, Raghavan, Mischler, and
  Mesgarani}]{styletts2}
Yinghao~Aaron Li, Cong Han, Vinay Raghavan, Gavin Mischler, and Nima Mesgarani.
  2024.
\newblock Styletts 2: Towards human-level text-to-speech through style
  diffusion and adversarial training with large speech language models.
\newblock \emph{Advances in Neural Information Processing Systems}, 36.

\bibitem[{Naeem et~al.(2020)Naeem, Oh, Uh, Choi, and Yoo}]{pmlr-v119-naeem20a}
Muhammad~Ferjad Naeem, Seong~Joon Oh, Youngjung Uh, Yunjey Choi, and Jaejun
  Yoo. 2020.
\newblock \href {https://proceedings.mlr.press/v119/naeem20a.html} {Reliable
  fidelity and diversity metrics for generative models}.
\newblock In \emph{Proceedings of the 37th International Conference on Machine
  Learning}, volume 119 of \emph{Proceedings of Machine Learning Research},
  pages 7176--7185. PMLR.

\bibitem[{Nguyen et~al.(2023)Nguyen, Hsu, d'Avirro, Shi, Gat, Fazel-Zarani,
  Remez, Copet, Synnaeve, Hassid et~al.}]{nguyen2023expresso}
Tu~Anh Nguyen, Wei-Ning Hsu, Antony d'Avirro, Bowen Shi, Itai Gat, Maryam
  Fazel-Zarani, Tal Remez, Jade Copet, Gabriel Synnaeve, Michael Hassid, et~al.
  2023.
\newblock Expresso: A benchmark and analysis of discrete expressive speech
  resynthesis.
\newblock \emph{arXiv preprint arXiv:2308.05725}.

\bibitem[{Nguyen et~al.(2024)Nguyen, Muller, Yu, Costa-Jussa, Elbayad, Popuri,
  Duquenne, Algayres, Mavlyutov, Gat et~al.}]{nguyen2024spirit}
Tu~Anh Nguyen, Benjamin Muller, Bokai Yu, Marta~R Costa-Jussa, Maha Elbayad,
  Sravya Popuri, Paul-Ambroise Duquenne, Robin Algayres, Ruslan Mavlyutov, Itai
  Gat, et~al. 2024.
\newblock Spirit-lm: Interleaved spoken and written language model.
\newblock \emph{arXiv preprint arXiv:2402.05755}.

\bibitem[{Panayotov et~al.(2015)Panayotov, Chen, Povey, and
  Khudanpur}]{Panayotov2015}
Vassil Panayotov, Guoguo Chen, Daniel Povey, and Sanjeev Khudanpur. 2015.
\newblock {LibriSpeech}: an {ASR} corpus based on public domain audio books.
\newblock In \emph{ICASSP}.

\bibitem[{Polyak et~al.(2021{\natexlab{a}})Polyak, Adi, Copet, Kharitonov,
  Lakhotia, Hsu, Mohamed, and Dupoux}]{polyak21_interspeech}
Adam Polyak, Yossi Adi, Jade Copet, Eugene Kharitonov, Kushal Lakhotia,
  Wei-Ning Hsu, Abdelrahman Mohamed, and Emmanuel Dupoux. 2021{\natexlab{a}}.
\newblock {Speech Resynthesis from Discrete Disentangled Self-Supervised
  Representations}.
\newblock In \emph{Proc. Interspeech 2021}.

\bibitem[{Polyak et~al.(2021{\natexlab{b}})Polyak, Adi, Copet, Kharitonov,
  Lakhotia, Hsu, Mohamed, and Dupoux}]{Polyak2021}
Adam Polyak, Yossi Adi, Jade Copet, Eugene Kharitonov, Kushal Lakhotia,
  Wei-Ning Hsu, Abdelrahman Mohamed, and Emmanuel Dupoux. 2021{\natexlab{b}}.
\newblock {Speech Resynthesis from Discrete Disentangled Self-Supervised
  Representations}.
\newblock In \emph{Interspeech}.

\bibitem[{Pratap et~al.(2020)Pratap, Xu, Sriram, Synnaeve, and Collobert}]{mls}
Vineel Pratap, Qiantong Xu, Anuroop Sriram, Gabriel Synnaeve, and Ronan
  Collobert. 2020.
\newblock {MLS}: A large-scale multilingual dataset for speech research.
\newblock \emph{arXiv preprint arXiv:2012.03411}.

\bibitem[{Radford et~al.(2021)Radford, Kim, Hallacy, Ramesh, Goh, Agarwal,
  Sastry, Askell, Mishkin, Clark et~al.}]{radford2021learning}
Alec Radford, Jong~Wook Kim, Chris Hallacy, Aditya Ramesh, Gabriel Goh,
  Sandhini Agarwal, Girish Sastry, Amanda Askell, Pamela Mishkin, Jack Clark,
  et~al. 2021.
\newblock Learning transferable visual models from natural language
  supervision.
\newblock In \emph{International conference on machine learning}, pages
  8748--8763. PMLR.

\bibitem[{Ren et~al.(2020)Ren, Hu, Tan, Qin, Zhao, Zhao, and Liu}]{Ren2020}
Yi~Ren, Chenxu Hu, Xu~Tan, Tao Qin, Sheng Zhao, Zhou Zhao, and Tie-Yan Liu.
  2020.
\newblock {FastSpeech 2}: Fast and high-quality end-to-end text to speech.
\newblock In \emph{ICLR}.

\bibitem[{Rubenstein et~al.(2023)Rubenstein, Asawaroengchai, Nguyen, Bapna,
  Borsos, de~Chaumont~Quitry, Chen, Badawy, Han, Kharitonov, Muckenhirn,
  Padfield, Qin, Rozenberg, Sainath, Schalkwyk, Sharifi, Tadmor, Ramanovich,
  Tagliasacchi, Tudor, Velimirović, Vincent, Yu, Wang, Zayats, Zeghidour,
  Zhang, Zhang, Zilka, and Frank}]{audiopalm}
Paul~K. Rubenstein, Chulayuth Asawaroengchai, Duc~Dung Nguyen, Ankur Bapna,
  Zalán Borsos, Félix de~Chaumont~Quitry, Peter Chen, Dalia~El Badawy, Wei
  Han, Eugene Kharitonov, Hannah Muckenhirn, Dirk Padfield, James Qin, Danny
  Rozenberg, Tara Sainath, Johan Schalkwyk, Matt Sharifi, Michelle Tadmor,
  Ramanovich, Marco Tagliasacchi, Alexandru Tudor, Mihajlo Velimirović, Damien
  Vincent, Jiahui Yu, Yongqiang Wang, Vicky Zayats, Neil Zeghidour, Yu~Zhang,
  Zhishuai Zhang, Lukas Zilka, and Christian Frank. 2023.
\newblock {AudioPaLM}: A large language model that can speak and listen.

\bibitem[{Saeed et~al.(2021)Saeed, Grangier, and
  Zeghidour}]{contrastive_learning_audio_rep}
Aaqib Saeed, David Grangier, and Neil Zeghidour. 2021.
\newblock \href {https://doi.org/10.1109/ICASSP39728.2021.9413528} {Contrastive
  learning of general-purpose audio representations}.
\newblock In \emph{ICASSP 2021 - 2021 IEEE International Conference on
  Acoustics, Speech and Signal Processing (ICASSP)}, pages 3875--3879.

\bibitem[{Sajjadi et~al.(2018)Sajjadi, Bachem, Lucic, Bousquet, and
  Gelly}]{NEURIPS2018_f7696a9b}
Mehdi S.~M. Sajjadi, Olivier Bachem, Mario Lucic, Olivier Bousquet, and Sylvain
  Gelly. 2018.
\newblock \href
  {https://proceedings.neurips.cc/paper_files/paper/2018/file/f7696a9b362ac5a51c3dc8f098b73923-Paper.pdf}
  {Assessing generative models via precision and recall}.
\newblock In \emph{Advances in Neural Information Processing Systems},
  volume~31. Curran Associates, Inc.

\bibitem[{Schneider et~al.(2019)Schneider, Baevski, Collobert, and
  Auli}]{wav2vec}
Steffen Schneider, Alexei Baevski, Ronan Collobert, and Michael Auli. 2019.
\newblock \href {https://doi.org/10.21437/Interspeech.2019-1873} {{wav2vec:
  Unsupervised Pre-Training for Speech Recognition}}.
\newblock In \emph{Proc. Interspeech 2019}, pages 3465--3469.

\bibitem[{Schroff et~al.(2015)Schroff, Kalenichenko, and Philbin}]{tripletloss}
Florian Schroff, Dmitry Kalenichenko, and James Philbin. 2015.
\newblock Facenet: A unified embedding for face recognition and clustering.
\newblock In \emph{Proceedings of the IEEE conference on computer vision and
  pattern recognition}, pages 815--823.

\bibitem[{Shen et~al.(2019)Shen, Ott, Auli, and Ranzato}]{pairwise_bleu}
Tianxiao Shen, Myle Ott, Michael Auli, and Marc'Aurelio Ranzato. 2019.
\newblock \href {https://proceedings.mlr.press/v97/shen19c.html} {Mixture
  models for diverse machine translation: Tricks of the trade}.
\newblock In \emph{Proceedings of the 36th International Conference on Machine
  Learning}, volume~97 of \emph{Proceedings of Machine Learning Research},
  pages 5719--5728. PMLR.

\bibitem[{Shor et~al.(2020)Shor, Jansen, Maor, Lang, Tuval, Quitry,
  Tagliasacchi, Shavitt, Emanuel, and Haviv}]{shor2020towards}
Joel Shor, Aren Jansen, Ronnie Maor, Oran Lang, Omry Tuval, Felix de~Chaumont
  Quitry, Marco Tagliasacchi, Ira Shavitt, Dotan Emanuel, and Yinnon Haviv.
  2020.
\newblock Towards learning a universal non-semantic representation of speech.
\newblock \emph{arXiv preprint arXiv:2002.12764}.

\bibitem[{Shu et~al.(2019)Shu, Nakayama, and Cho}]{shu-etal-2019-generating}
Raphael Shu, Hideki Nakayama, and Kyunghyun Cho. 2019.
\newblock \href {https://doi.org/10.18653/v1/P19-1177} {Generating diverse
  translations with sentence codes}.
\newblock In \emph{Proceedings of the 57th Annual Meeting of the Association
  for Computational Linguistics}, pages 1823--1827, Florence, Italy.
  Association for Computational Linguistics.

\bibitem[{Spearman(1904)}]{spearman04}
C.~Spearman. 1904.
\newblock The proof and measurement of association between two things.
\newblock \emph{American Journal of Psychology}, 15:88--103.

\bibitem[{Stoidis and Cavallaro(2022)}]{stoidis22_interspeech}
Dimitrios Stoidis and Andrea Cavallaro. 2022.
\newblock \href {https://doi.org/10.21437/Interspeech.2022-11322} {Generating
  gender-ambiguous voices for privacy-preserving speech recognition}.
\newblock In \emph{Interspeech 2022}, pages 4237--4241.

\bibitem[{Tevet and Berant(2021)}]{tevet-berant-2021-evaluating}
Guy Tevet and Jonathan Berant. 2021.
\newblock \href {https://doi.org/10.18653/v1/2021.eacl-main.25} {Evaluating the
  evaluation of diversity in natural language generation}.
\newblock In \emph{Proceedings of the 16th Conference of the European Chapter
  of the Association for Computational Linguistics: Main Volume}, pages
  326--346, Online. Association for Computational Linguistics.

\bibitem[{van~den Oord et~al.(2018)van~den Oord, Li, and Vinyals}]{cpc}
A{\"{a}}ron van~den Oord, Yazhe Li, and Oriol Vinyals. 2018.
\newblock Representation learning with contrastive predictive coding.
\newblock \emph{arXiv:1807.03748}.

\bibitem[{Veaux et~al.(2016)Veaux, Yamagishi, and MacDonald}]{vctk}
Christophe Veaux, Junichi Yamagishi, and Kirsten MacDonald. 2016.
\newblock \href {https://doi.org/10.7488/ds/1495} {Cstr vctk corpus: English
  multi-speaker corpus for cstr voice cloning toolkit}.

\bibitem[{Vyas et~al.(2023)Vyas, Shi, Le, Tjandra, Wu, Guo, Zhang, Zhang,
  Adkins, Ngan, Wang, Cruz, Akula, Akinyemi, Ellis, Moritz, Yungster,
  Rakotoarison, Tan, Summers, Wood, Lane, Williamson, and Hsu}]{audiobox}
Apoorv Vyas, Bowen Shi, Matthew Le, Andros Tjandra, Yi-Chiao Wu, Baishan Guo,
  Jiemin Zhang, Xinyue Zhang, Robert Adkins, William Ngan, Jeff Wang, Ivan
  Cruz, Bapi Akula, Akinniyi Akinyemi, Brian Ellis, Rashel Moritz, Yael
  Yungster, Alice Rakotoarison, Liang Tan, Chris Summers, Carleigh Wood, Joshua
  Lane, Mary Williamson, and Wei-Ning Hsu. 2023.
\newblock \href {https://arxiv.org/abs/2312.15821} {Audiobox: Unified audio
  generation with natural language prompts}.
\newblock \emph{Preprint}, arXiv:2312.15821.

\bibitem[{Wang et~al.(2023)Wang, Chen, Wu, Zhang, Zhou, Liu, Chen, Liu, Wang,
  Li et~al.}]{valle}
Chengyi Wang, Sanyuan Chen, Yu~Wu, Ziqiang Zhang, Long Zhou, Shujie Liu, Zhuo
  Chen, Yanqing Liu, Huaming Wang, Jinyu Li, et~al. 2023.
\newblock Neural codec language models are zero-shot text to speech
  synthesizers.
\newblock \emph{arXiv preprint arXiv:2301.02111}.

\bibitem[{Zeghidour et~al.(2021)Zeghidour, Luebs, Omran, Skoglund, and
  Tagliasacchi}]{soundstream}
Neil Zeghidour, Alejandro Luebs, Ahmed Omran, Jan Skoglund, and Marco
  Tagliasacchi. 2021.
\newblock {SoundStream}: An end-to-end neural audio codec.
\newblock \emph{IEEE/ACM Transactions on Audio, Speech, and Language
  Processing}.

\bibitem[{Zen et~al.(2019)Zen, Dang, Clark, Zhang, Weiss, Jia, Chen, and
  Wu}]{zen2019libritts}
Heiga Zen, Viet Dang, Rob Clark, Yu~Zhang, Ron~J Weiss, Ye~Jia, Zhifeng Chen,
  and Yonghui Wu. 2019.
\newblock Libritts: A corpus derived from librispeech for text-to-speech.

\bibitem[{Zhang et~al.(2020)Zhang, Kishore, Wu, Weinberger, and
  Artzi}]{Zhang*2020BERTScore:}
Tianyi Zhang, Varsha Kishore, Felix Wu, Kilian~Q. Weinberger, and Yoav Artzi.
  2020.
\newblock \href {https://openreview.net/forum?id=SkeHuCVFDr} {Bertscore:
  Evaluating text generation with bert}.
\newblock In \emph{International Conference on Learning Representations}.

\bibitem[{Zhu et~al.(2018)Zhu, Lu, Zheng, Guo, Zhang, Wang, and Yu}]{selfbleu}
Yaoming Zhu, Sidi Lu, Lei Zheng, Jiaxian Guo, Weinan Zhang, Jun Wang, and Yong
  Yu. 2018.
\newblock Texygen: A benchmarking platform for text generation models.
\newblock In \emph{The 41st international ACM SIGIR conference on research \&
  development in information retrieval}, pages 1097--1100.

\end{thebibliography}
\bibliographystyle{acl_natbib}

\appendix

\section{Training Data}\label{ss:data}
To train SpeechSim, we use LibriLight \citep{Kahn2020}, an unlabelled dataset containing approximately $ 60,000$ hours of self-recorded audiobooks. To train projection models on top of SpeechSim, we use different datasets depending on the target facet. Specifically, the projection models related to speaker voices and gender are trained on LibriTTS train-other~\citep{zen2019libritts}. The projector related to emotion is trained on EmoV \citep{adigwe2018emotional} and Expresso \citep{nguyen2023expresso} train splits. For Expresso, we selected audios with a single speaker and removed narration and enunciated classes. Expresso and EmoV datasets have only the neutral class in common. 

Next, the accent projection model is trained on VCTK-train~\citep{vctk}, which contains 12 English accents spoken by 110 speakers. Finally, the background noise projection model is trained on AudioSet \citep{Gemmeke2017AudioSA}. Here, we select all examples that are tagged as ``Speech'' and which have between 1 and 3 additional tags as we found examples with more tags to be too noisy. We randomly sample 70\% of the selected examples to build the train set, the remaining 30\% are randomly split as dev and test.

\section{Baseline Representations}\label{ss:baseline_representations}

\subsection{Generic representations}

\paragraph{Wav2Vec-BERT} We experiment with representations provided by a 600M-parameter Wav2Vec-BERT model~\cite{w2vb}. We take embeddings of intermediate layers, with each utterance represented as a sequence of 1024-dimensional vectors corresponding to 40ms frame.

\paragraph{HuBERT} We use HuBERT embeddings, produced by the Base model~\cite{hubert}  (95M parameters), trained on LibriSpeech~\cite{Panayotov2015}. We follow~\citet{Lakhotia2021} and use embeddings extracted from the 6th layer. This model has a framerate of 20 ms.

\paragraph{SoundStream} Another type of representations found to be useful in AudioLM is SoundStream~\cite{soundstream}. While Wav2Vec-BERT representations were found to be high-level, SoundStream representations are fine-grained, faithfully representing minute details of speech~\cite{Borsos2023}. We use a  version of SoundStream with 60 residual quantization layers, each quantized into 1024 tokens and 200 ms frames. The encoder has around 12M parameters.
For uniformity, we use dequantized representations. 
This model operates on 24kHz audio.

\subsection{Specialized representations} 
\paragraph{Training from scratch} To evaluate the contribution of the pretrained model SpeechSim, we also trained a model from scratch for each facet on the labelled data directly. These models have the same architecture as SpeechSim but with 6 layers. They are trained with the same contrastive objective used for training the projection models.

\section{Effect of decoding method and temperature on acoustic diversity}\label{ss:effect-diversity}
\subsection{SPEAR-TTS: Effect of Best-of-K decoding}
\label{ss:bestk}
\citet{spear-tts} leveraged the stochasticity of the temperature decoding to improve audio quality of the generated speech in an AudioLM-like model. Namely, they proposed to sample $K$ examples for the same input and then select the sample with the highest acoustic quality as the output. Now we check if that approach brings a penalty to the acoustic diversity.

Since SPEAR-TTS is only stochastic in the semantic-to-acoustic token decoding, we use the same task and AudioLM model as above. Again, for each semantic token sequence, we sample 128 acoustic token sequences that correspond to it. Next, we randomly split them in groups of $K \in \{1, 2, 4\}$ and, within each group, select one with the highest acoustic quality score, relying on the same model as \citet{spear-tts}. After this filtering, each semantic token sequence is mapped into 128, 64, or 32 audio samples (for K=1, 2, and 4). In the first two cases, we sample 32 sequences so that all sets have the same size.

From Table~\ref{tab:best_of_k_wins} see that, except for the gender facet, higher $K$ always brings a diversity penalty.

\begin{table}
\centering
\resizebox{\linewidth}{!}{
\begin{tabular}{lcccccc}
\toprule
K & Voice & Gender & Emotion & Accent & Background Noise \\
\toprule
1 & 64 & 41 & 62 & 63 & 61 \\
4 & 0 & 27 & 0 & 0 & 0 & \\
\bottomrule
\end{tabular}
}
\caption{Best-of-K filtering on the semantic-to-acoustic tokens resynthesis task. The proportion (\%) of utterances where Best-K with $K \in \{1, 4\}$ have higher diversity scores than with $K=2$.}
\label{tab:best_of_k_wins}
\end{table}

\subsection{Text Synthesis: Influence of Temperature}\label{ss:temp}
Temperature is a parameter aiming at controlling the sampling distribution. As the sampling temperature approaches zero, samples for a fixed input become increasingly deterministic; while a high temperature introduces diversity in the sampling process. A priori, one can expect that this token-level diversity can translate to higher-level, per-utterance acoustic diversity.
In this experiment we use a single-stage fully auto-regressive TTS system that maps grapheme text representations to a sequence of acoustic tokens, akin to VALL-E~\cite{valle}. This model is trained on a combination of Multilingual LibriSpeech~\cite{mls}, LibriLight~\cite{Kahn2020} and other public data. Table~\ref{tab:temp} shows that voice diversity is by far the most affected by small variations of temperature while gender and background noise are barely impacted. Samples with low temperature are also less diverse in terms of emotion and accent.

\begin{table}[tb]
\centering
\resizebox{\linewidth}{!}{
\begin{tabular}{lccccccccc}
\toprule
T & Voice & Gender & Emotion & Accent & Background Noise \\
\midrule
0.7 & 12 & 33 & 21 & 18 & 29  \\
0.9 & 45 & 21 & 40 & 40 & 36 \\
\bottomrule
\end{tabular}
}
\caption{Count (out of 64) of instances where sets of utterances generated with $T \in \{0.7, 0.9\}$ have higher diversity scores than with  $T=0.8$.}
\label{tab:temp}
\end{table}

\section{Additional experiments on testing whether facets are mixed up together} \label{ss:facets-mix}

\subsection{Emotions and Voices} We examine the interactions between emotion and voice diversity within metrics as explained in Section~\ref{ss:entangled}. The number of speakers is 1 or 4 depending on the diversity level of the set of utterances. For sets with high entropy (i.e. high diversity), there is an unique speaker voice while there are 4 speakers for sets with low entropy (i.e. low diversity). We report the results in Table~\ref{tab:emotion_control_speakers_tab}. We observe that SpeechSim/Emotion is unaffected by the changes in the number of speakers, while SpeechSim/Voice follows the speaker diversity and hence it has a negative correlation with the emotion diversity score. SpeechSim focuses on the changes in the speaker diversity, thus having a negative correlation with the gender diversity.

\subsection{Varying accents and voices in opposite directions} \label{ss:accent-mix}
In this experiment, we use 7 voices for sets with high entropy and 30 voices for sets with low entropy. The results are reported in Table~\ref{tab:accent_control_speakers_tab}. We see that the projected metric is unaffected by changes in voices, while the vanilla version of SpeechSim is focused on the speaker facet.

\begin{table}[tb]
\centering
\resizebox{\linewidth}{!}{\begin{tabular}{lcc}
\toprule
& \multicolumn{2}{c}{EmoV} \\
&    Avg. cosine               &       Vendi score  \\ \midrule

SpeechSim & -0.384 (± 0.347)  &  -0.372 (± 0.347)     \\

SpeechSim/Voice & -0.609 (± 0.185)  &  -0.451 (± 0.287)  \\

SpeechSim/Emotion &  \hphantom{-}0.970 (± 0.042)  &  \hphantom{-}{0.985} (± 0.028)    \\

\midrule

& \multicolumn{2}{c}{Expresso} \\
     &    Avg. cosine               &       Vendi score  \\ \midrule
SpeechSim & -0.257 (± 0.423) &  -0.218 (± 0.423)    \\
SpeechSim/Voice & -0.397 (± 0.464) & -0.290 (± 0.530)  \\
SpeechSim/Emotion &   \hphantom{-}{0.994} (± 0.013) &  \hphantom{-}{0.985} (± 0.019)  \\

\bottomrule
\end{tabular}}
\caption{Changing emotion and speaker diversity in the opposite directions. Average Spearman correlation between emotion diversity and diversity scores induced by speech representations.}
\label{tab:emotion_control_speakers_tab}
\end{table}

\begin{table}[tb]
\centering
\resizebox{\linewidth}{!}{\begin{tabular}{lcc}
\toprule

&    Avg. cosine               &       Vendi score  \\ \midrule

SpeechSim & -0.677 (± 0.221)  &  -0.737 (± 0.119)     \\


SpeechSim/Voice & -0.743 (± 0.199)  &  -0.790 (± 0.110)  \\ 
SpeechSim/Accent &  \hphantom{-}0.795 (± 0.159)  &  \hphantom{-}{0.999} (± 0.008)   \\

\bottomrule
\end{tabular}}
\caption{Changing accent and voice diversity in the opposite directions. Average Spearman correlation between accent diversity and diversity scores induced by speech representations.}
\label{tab:accent_control_speakers_tab}
\end{table}

\section{Acoustic diversity in Text-to-Speech systems - Model details} \label{ss:tts-models}

\textit{Bark TTS}\footnote{\url{https://github.com/suno-ai/bark}} is a generative text-to-audio system that is based on  a decoder-only Transformer model that operates on discretized speech representations, akin to AudioLM~\cite{Borsos2023} and VALL-E~\cite{valle}. Bark TTS synthesizes utterances in a random voice and acoustic conditions.

\textit{StyleTTS 2} \citep{styletts2} combines a style diffusion model with speech with WavLM-based adversarial training~\cite{wavlm} for an end-to-end training of a multi-component system. We use a checkpoint trained on LibriTTS~\cite{zen2019libritts}. The samples are conditioned on a female voice, but randomize timbre and prosody.\footnote{\texttt{\href{https://colab.research.google.com/github/yl4579/StyleTTS2/blob/main/Demo/Inference_LibriTTS.ipynb}{https://colab.../Inference\_LibriTTS.ipynb}}}

\textit{Tortoise TTS} \citep{betker2023better} combines auto-regressive and diffusion-based components~\cite{tortoisetts}. For training, it pooled LibriTTS~\cite{zen2019libritts}, HiFiTTS~\cite{bakhturina2021}, and public data sourced from audiobooks and podcasts.

\textit{FastSpeech 2} \citep{Ren2020} consists of a Transformer-based encoder that embeds phoneme-based text representation, a variance adapter that predicts pitch, energy, and duration values, and a non-autoregressive decoder. We use a publicly available re-implementation trained on LJSpeech~\cite{ljspeech}.\footnote{\url{https://github.com/ming024/FastSpeech2}}

\end{document}